\documentclass{aastex701}

\newcommand{\included}{\checkmark}

\shorttitle{Processed Dust Distribution in Evolving Disks}
\shortauthors{Ishizaki et al.}
\submitjournal{ApJ}
\graphicspath{{./}{figures/}}
\begin{document}

\title{Distribution of Chemically-Processed Dust in a Viscously Evolving Protoplanetary Disk:
  Application to Crystalline Silicates in Comets}

\author[orcid=0009-0005-8583-9730,gname=Lily,sname=Ishizaki]{Lily Ishizaki}
\affiliation{Department of Earth and Planetary Science, The University of Tokyo, Hongo, Tokyo 113-0033, Japan}
\email[show]{r.ishizaki@eps.s.u-tokyo.ac.jp}  

\author[orcid=0000-0002-4603-9440,gname=Shogo, sname=Tachibana]{Shogo Tachibana} 
\affiliation{Department of Earth and Planetary Science, The University of Tokyo, Hongo, Tokyo 113-0033, Japan}
\email{tachi@eps.s.u-tokyo.ac.jp}

\author[gname=Shigeru, sname=Ida]{Shigeru Ida} 
\affiliation{Department of Astronomy, School of Science, Westlake University, Hangzhou, Zhejiang 310030, China}
\affiliation{Earth-Life Science Institute, Institute of Science Tokyo, Tokyo 152-8550, Japan}
\email{ida@westlake.edu.cn}

\begin{abstract}
  Dust particles undergo chemical reactions in protoplanetary disks according to their environments,
  producing compositional diversity in planetary materials.
  Extraterrestrial records of irreversible reactions, such as crystallization of amorphous silicates,
  provide particularly strong constraints on the early evolution of the protosolar disk.
  In this study, we investigate such irreversible reactions and the spatiotemporal distribution
  of reacted dust in a viscously evolving disk using Monte Carlo particle-tracking simulations.
  We extend a predictive formula for the temperature at which irreversible reactions proceed efficiently (“reaction line”),
  originally developed for steady accretion disks, to viscously expanding disks.
  The spatiotemporal distribution of reacted dust is governed by the relative locations of the reaction line
  and the stagnation line, which separates inward and outward advection in the disk.
  The reaction line moves inward as the disk cools,
  while the stagnation line moves outward owing to the radial viscous spreading of the disk.
  When the reaction line lies far inside the stagnation line, the reacted dust remains inside the reaction line.
  On the other hand, when the reaction line lies near or beyond the stagnation line,
  the reacted dust located near the stagnation line or between the two lines is transported outward efficiently.
  It results in a radially broad distribution of reacted dust throughout the disk,
  including the outer regions where the temperatures remain too low for reactions.
  We assessed the disk conditions consistent with the crystalline silicates observed in Solar System comets
  and found that the protosolar disk was likely compact, moderately massive, and not strongly turbulent.
\end{abstract}

\section{Introduction} \label{sec:introduction}
Dust particles undergo various chemical reactions in protoplanetary disks,
producing chemical diversity in the planetary materials formed from them.
The reaction histories recorded in Solar System materials provide critical constraints
on the physical conditions that governed the early Solar System.
In particular, irreversible dust reactions offer especially strong constraints on the protosolar disk,
because once they proceed, dust particles permanently record the conditions they experienced.

Numerous signatures of such reactions have been identified in extraterrestrial materials and astronomical observations.
Crystalline silicates, formed by annealing of amorphous silicates, have been detected
in cometary samples originating from the cold outer disk \citep{Brownlee2006Science, Zolensky2006Science},
and 10-60$\%$ of cometary silicates are crystalline
\citep{Harker2011TheAstronomicalJournal, Sitko2011TheAstronomicalJournal, Shinnaka2018AJ},
implying outward transport from the hot inner disk.
Comet-to-comet differences in the olivine-to-pyroxene ratio \citep{Shinnaka2018AJ} further suggest varying degrees of outward transport
and dynamical mixing following high-temperature processing, because crystallization of amorphous pyroxene and amorphous olivine requires
different heating conditions \citep{Yamamoto2018ACSEarthSpaceChem., Kobayashi2023}.

\citet{Ciesla2011ApJ} demonstrated that turbulent diffusion can transport crystalline silicates outward over large radial distances.
However, they primarily focused on particle transport and thermal histories,
and their model was not designed to quantitatively reproduce the observed crystalline abundances in comet forming regions.
\citet{Okamoto2022ApJ} has incorporated the pebble accretion into a simple steady-accretion disk,
in which silicate grains are embedded in icy pebbles and are released at the snowline.
They showed that turbulent diffusion alone produces a crystalline fraction of only $\sim$5\%,
and that additional processes, such as the sticking of silicate particles onto drifting icy pebbles
and the decay of pebble flux, are required to match the observed abundances.

These results imply that reproducing the observed crystalline abundances requires a framework that directly links
dust transport and chemical processing to the evolving disk model.
Moreover, the extraterrestrial materials we analyze today do not represent snapshots of intermediate stages,
but rather the final outcomes of time dependent physical and chemical evolution.
Interpreting such records within a time-independent disk model, as adopted by \citet{Okamoto2022ApJ},
is therefore insufficient to fully capture their evolutionary origins.
Instead, reconstructing their origins requires tracing these histories backward from their end states.
To achieve this, it is necessary to model how the disk itself evolved,
and how the associated chemical environments influenced the dust particles.

In this study, we investigate the transport and chemical evolution of dust particles in a viscously evolving protoplanetary disk.
We develop a viscously evolving protoplanetary disk model that connects a viscous-heating solution in the inner disk
with a self-similar irradiative-heating solution in the outer disk, conserving disk accretion flux \citep{Chambers2009ApJ,Ida2016A&A},
which we refer to as a combined ``viscous-irradiative'' disk model.
This framework reproduces viscous heating in the hot inner region, where dust chemistry proceeds,
while allowing outward mass transport through disk expansion.
By incorporating experimentally determined reaction kinetics into an evolving disk model,
we track the thermal histories of individual dust particles and evaluate their reaction progress
during transport driven by advection and diffusion.
We then apply this framework to examine whether the crystalline abundances observed
in comets can be reproduced through disk evolution alone,
without invoking additional processes such as dust sticking onto icy pebbles.

In Section~\ref{sec:method}, we construct a combined viscous-irradiative disk model,
which self-consistently includes both viscous heating and stellar irradiation,
and describe a 3D Monte Carlo method to track the trajectories of dust particles and their chemical evolution.
Section~\ref{sec:results} analyzes the distribution of dust that undergoes various irreversible reactions.
We further extend the prediction framework of the “reaction line,” originally developed for steady accretion disks,
to viscously evolving disks and classify the spatial distributions of reacted dust
based on the interplay between reaction progress and particle transport.
We will show that the relative configurations of the reaction line and the disk accretion reversal radius (``stagnation line")
is a crucial factor for the final radial distribution of the reacted particles.
Finally, we apply the model to crystallization in the early Solar System
and summarize the conclusions in Section~\ref{sec:conclusions}.

\section{Method} \label{sec:method}
We constructed a time-evolving protoplanetary disk model that includes both viscous and stellar-irradiative heating
by connecting a steady accretion solution in the inner disk with a self-similar solution in the outer disk.
Previous hybrid disk models \citep{Chambers2009ApJ,Ida2016A&A} were primarily developed to investigate
the physical evolution of solid materials in large disks and employed thermal structures appropriate for those applications.
In this study, we focus on chemical reactions of dust particles that proceed in the hot inner disk,
where the local thermal structure is strongly affected by viscous dissipation.
Following \citet{Ishizaki2023ApJ}, we assume that viscous heating is distributed in proportion to the local gas density
and derive the disk temperature structure from local viscous heating and stellar irradiation.
In addition, we include a correction for the exponential tail region of the disk,
which becomes important when applying the model to initally compact disks.

In this study, the mass and luminosity of the central star are set to solar values,
as we focus specifically on modeling the Solar System. The disk gas
is assumed to be a mixture of $\mathrm{H_2}$ and $\mathrm{He}$, with a fixed mean molecular weight
of $\mu = 2.34\,\mathrm{g\,mol^{-1}}$. The opacity $\kappa$
is also fixed at $2.5\,\mathrm{cm^{2}\,g^{-1}}$ \citep{Pollack1985Icarus,Pollack1994ApJ}.
For simplicity, the functional dependence of derived quantities (e.g., $T(t,r,z)$, $\Sigma_{\mathrm{g}}(t,r)$)
on $\mu$ and $\kappa$ is not considered in the following discussion.
Note that these parameters are fixed in this study, but can be varied within the framework of the developed model.

We adopted the $\alpha$-viscosity model \citep{Shakura1973Astron.Astrophys.}
to account for viscous heating and turbulent transport in the disk:
The gas turbulent viscosity is given by $\nu = \alpha H c_{s}$, where $\alpha$ is a dimensionless viscous parameter,
$H$ is the local gas scale height, and $c_{s}$ is the local speed of sound.
Here, the scale height is defined as $H = c_{s}/\Omega_{K}$, with $\Omega_{K}$
being the local Keplerian frequency ($\Omega_{K} = \sqrt{G M_* / r^3}$;
$M_*$ is the mass of the central star).

The combined viscous-irradiative disk model consists of two regions:
an inner region dominated by viscous heating and an outer region dominated by stellar irradiation \citep{Chambers2009ApJ, Ida2016A&A}.
In the outer region, the disk temperature under irradiation is given by \citep{Kusaka1970Prog.Theor.Phys.}

\begin{equation}
  T_{\mathrm{irr}}=130\left(\frac{r}{1 \mathrm{au}}\right)^{-\frac{3}{7}} \mathrm{K},
\end{equation}
independently of vertical height $z$. This temperature profile corresponds
to a kinematic viscosity that scales as $\nu \propto r^{15/14}$,
which yields the following self-similar solution for the surface density
\citep{Lynden-Bell1974Mon.Not.R.Astron.Soc., Hartmann1998ApJ}:
\begin{equation}
  \Sigma_{\mathrm{g,irr}}=\Sigma_{\mathrm{g} 0}\left(\frac{t}{t_{\mathrm{diff}}}+1\right)^{-\frac{3 / 2-\gamma}{2-\gamma}}
  \left(\frac{r}{r_{\mathrm{d0}}}\right)^{-\gamma} \exp \left(-\frac{\left(r / r_{\mathrm{d} 0}\right)^{2-\gamma}}
    {\left(t / t_{\mathrm{diff}}\right)+1}\right),
\end{equation}
where $\gamma = 15/14$, $\Sigma_{\mathrm{g0}}$ is the initial surface density
at $r=r_{\mathrm{d0}}$, and $t_{\mathrm{diff}}$ is the characteristic diffusion timescale,
given by
\begin{equation}
  t_{\mathrm{diff}}=\left.\frac{1}{3(2-\gamma)^2} \frac{r^2}{\nu}\right|_{r=r_{\mathrm{d} 0}}
\end{equation}
Since $\gamma \sim 1$, the above expressions can be significantly simplified
by assuming $\nu \propto r$ which allows for analytical treatment \citep{Ida2016A&A}.
In this simplified form, the surface density becomes
\begin{equation}
  \Sigma_{\mathrm{g,irr}}=\Sigma_{\mathrm{g} 0}\left(\frac{t}{t_{\mathrm{diff}}}+1\right)^{-\frac{3}{2}}
  \left(\frac{r}{r_{\mathrm{d} 0}}\right)^{-1} \exp \left(-\frac{r}{r_{\mathrm{d}}(t)}\right). 
  \label{eq:sigma_general}
\end{equation}
with the diffusion timescale now defined as
\begin{equation}
  t_{\mathrm{diff}}=\left.\frac{r^2}{3 \nu}\right|_{r=r_{\mathrm{d} 0}},
\end{equation}
and the time-dependent scale radius evolves $r_{\mathrm{d}}(t)$ as
\begin{equation}
  r_{\mathrm{d}}(t)=\left(\frac{t}{t_{\mathrm{diff}}}+1\right) r_{\mathrm{d} 0}
\end{equation}
Here, $r_{\mathrm{d}}(t)$ corresponds to the exponential cutoff radius that characterizes the scale
over which the surface density declines, and changes with time.
In this paper, we adopt a broad range of $r_{\mathrm{d} 0}$ in 1-50 au.
As we will show later, a relatively small value of $r_{\mathrm{d} 0}$ is essential
to bring crystallized particles to the disk outer regions.

For consistency with this simplified structure, the irradiation temperature profile is also approximated by
\begin{equation}
  T_{\mathrm{irr}}=130\left(\frac{r}{1 \mathrm{au}}\right)^{-\frac{1}{2}} \mathrm{K},
  \label{eq:temperature_irr}
\end{equation}
which corresponds to a disk scale height of
\begin{equation}
  H_{\mathrm{irr}}=0.0221\left(\frac{r}{1 \mathrm{au}}\right)^{\frac{5}{4}} \mathrm{au}.
\end{equation}

The radial drift velocity due to gas drag is given by
\citep{Lynden-Bell1974Mon.Not.R.Astron.Soc., Hartmann1998ApJ}
\begin{equation}
  v_r=-\frac{3 \nu}{2 r}\left[1-\frac{r}{r_{\mathrm{d}}(t) / 2}\right]
  \label{eq:velo_r}
\end{equation}
and the dependence of $\dot{M}(r,t)$ on radius and time is
\begin{equation}
  \dot{M}(r,t)=-2 \pi r \Sigma_{\mathrm{g}} v_r=3 \pi \Sigma_{\mathrm{g}} \nu
  \left[1-\left(\frac{r}{r_{\mathrm{d}}(t) / 2}\right)\right].
\end{equation}
Since $3\pi \Sigma_{\mathrm{g}} \nu \propto \exp \left[-r/r_{\mathrm{d}}(t)\right]$,
it becomes independent of $r$ for $r \ll r_{\mathrm{d}}$.
Defining this constant as $\dot{M}_{\mathrm{in}}(t)$, we obtain
\begin{equation}
  \dot{M}(r,t)=\dot{M}_{\mathrm{in}}(t)\left[1-\left(\frac{r}{r_{\mathrm{d}}(t) / 2}\right)\right]
  \exp \left(-\frac{r}{r_{\mathrm{d}}(t)}\right),
  \label{eq:mdot}
\end{equation}
\begin{equation}
  \Sigma_{\mathrm{g}}=\frac{\dot{M}_{\mathrm{in}}(t)}{3 \pi \nu} \exp \left(-\frac{r}{r_{\mathrm{d}}(t)}\right).
  \label{eq:sigma_mdotin}
\end{equation}
The exponential cutoff term $\exp(-r/r_{\mathrm{d}}(t))$ appearing in Eqs.~(\ref{eq:mdot})
and (\ref{eq:sigma_mdotin}) originates from Eq.~(\ref{eq:sigma_general}).

Integrating Eq.~(\ref{eq:sigma_mdotin}) with respect to $r$, the disk mass
for the self-similar solution model $M_{\mathrm{d}}(t)$ is
\begin{equation}
  M_{\mathrm{d}}(t)=\int_0^{\infty} 2 \pi r \Sigma_{\mathrm{g}} \mathrm{dr}
  \approx 2 \pi \Sigma_{\mathrm{g} 0} r_{\mathrm{d} 0}^2\left(\frac{t}{t_{\mathrm{diff}}}+1\right)^{-\frac{1}{2}}
  \label{eq:mdself}
\end{equation}
Although the actual integrated mass of the disk is smaller than $M_{\mathrm{d}}(t)$
due to the presence of the inner viscous region as discussed below,
we adopt $M_{\mathrm{d}}(t)$ as a representative total disk mass,
since the outer irradiated region dominates the overall mass budget.
By substituting $\Sigma_{\mathrm{g0}}=M_{\mathrm{d} 0} / 2 \pi r_{\mathrm{d0}}^2$
($M_{\mathrm{d} 0}$ denotes the initial value of $M_{\mathrm{d}}(t)$) into Eq.~(\ref{eq:sigma_general}), we obtain
\begin{equation}
  \Sigma_{\mathrm{g}, \mathrm{irr}}
  =1.41 \times 10^3\left(\frac{r_{\mathrm{d} 0}}{30\,\mathrm{au}}\right)^{-1}\left(\frac{M_{\mathrm{d} 0}}
    {0.03 M_{\odot}}\right)\left(\frac{t}{t_{\mathrm{diff}}}+1\right)^{-\frac{3}{2}}
  \left(\frac{r}{1\,\mathrm{au}}\right)^{-1} \exp \left(-\frac{r}{r_{\mathrm{d}}(t)}\right) \mathrm{g} / \mathrm{cm}^2
  \label{eq:sigma_irr}
\end{equation}
where $M_{\odot}$ is the solar mass.

The relationship between $\dot{M}_{\mathrm{in}}(0)$ and $M_{\mathrm{d} 0}$ can be derived
from Eqs.~(\ref{eq:mdot})--(\ref{eq:sigma_irr}) as:
\begin{equation}
  \frac{\dot{M}_{\mathrm{in}}(0)}{10^{-8} M_{\odot} / \mathrm{yr}}=0.459\left(\frac{r_{\mathrm{d0}}}
    {30\,\mathrm{au}}\right)^{-1}\left(\frac{\alpha}{10^{-3}}\right)\left(\frac{M_{\mathrm{d} 0}}{0.03 M_{\odot}}\right).
\end{equation}

The inner region of the disk is assumed to be heated by uniform viscous heating,
proportional to the local gas spatial density \citep{Lin1985}.
A Gaussian density distribution of $\rho(z)=\left(\Sigma_g / \sqrt{2 \pi} H\right) \exp \left(-z^2 / 2 H^2\right)$
was assumed for vertical distribution of gas. Following \citet{Ishizaki2023ApJ},
the photosurface temperature of the inner disk is
\begin{equation}
  T_{\mathrm{vis,surf}} = 70
  \left(\frac{r_{\mathrm{d}0}}{30\,\mathrm{au}}\right)^{-\frac{1}{4}}
  \left(\frac{\alpha}{10^{-3}}\right)^{\frac{1}{4}}
  \left(\frac{M_{\mathrm{d} 0}}{0.03\,M_{\odot}}\right)^{\frac{1}{4}}
  \left(\frac{t}{t_{\mathrm{diff}}}+1\right)^{-\frac{3}{8}}
  \left(\frac{r}{1\,\mathrm{au}}\right)^{-\frac{3}{4}}
  \exp\left(-\frac{1}{4}\frac{r}{r_{\mathrm{d}}(t)}\right),
  \label{eq:temperature_surface}
\end{equation}
The exponential reduction term $\exp \left(-r / 4 r_{\mathrm{d}}(t)\right)$ in Eq.~(\ref{eq:temperature_surface})
originates from the decay behavior of the self-similar solution described in Eq.~(\ref{eq:sigma_general}),
combined with the relationship $T_{\mathrm{vis,surf}} \propto \Sigma^{1 / 4}$. Temperature at height $z$ is given by
\begin{equation}
  T_{\mathrm{vis}}(z)=\left[1+\frac{3\left(\tau_z-1\right)}{4}\left(1-\frac{\tau_z-1}{\tau_{\mathrm{d}}}
    \right)\right]^{\frac{1}{4}} T_{\mathrm{vis,surf }},
\end{equation}
where $\tau_{\mathrm{d}} $ is the disk optical depth, defined by $\tau_{\mathrm{d}} = \int_{-\infty}^{\infty} \kappa \rho(z)\, dz$.
The optical depth at $z$, $\tau_z$, is analytically obtained by
\begin{equation}
  \tau_z = \frac{\kappa \Sigma_{\mathrm{g}}}{2}
  \left[ 1 - \mathrm{erf} \left( \frac{z}{\sqrt{2}H} \right) \right],
\end{equation}
where $\mathrm{erf}(x)$ is the error function defined
by $\mathrm{erf}(x) = ( \sqrt{\pi} / 2 ) \int_{0}^{x} \exp(-x'^2)\, dx'$.
Because opacity $\kappa$ is constant in this model, temperature at the midplane $T_{\mathrm{vis,c}}$ is
\begin{equation}
  T_{\mathrm{vis,c}} = \left( \frac{3 \kappa \Sigma_{\mathrm{g}}}{16} \right)^{\frac{1}{4}} T_{\mathrm{vis,surf}},
\end{equation}
Considering $\Sigma_{\mathrm{g}} = \dot{M}_{\mathrm{in},0} \exp\left(-r_{\mathrm{d}}(t)/r\right)/3\pi \nu \propto
\dot{M}_{\mathrm{in},0}/T_{\mathrm{vis,c}} \exp\left(-{r_{\mathrm{d}}(t)/r}\right)$,
$T_{\mathrm{vis,c}}$ is derived as
\begin{equation}
  T_{\mathrm{vis,c}} = 288
  \left(\frac{r_{\mathrm{d}0}}{30\,\mathrm{au}}\right)^{-\frac{2}{5}}
  \left(\frac{\alpha}{10^{-3}}\right)^{\frac{1}{5}}
  \left(\frac{M_{\mathrm{d} 0}}{0.03\,M_{\odot}}\right)^{\frac{2}{5}}
  \left(\frac{t}{t_{\mathrm{diff}}}+1\right)^{-\frac{3}{5}}
  \left(\frac{r}{1\,\mathrm{au}}\right)^{-\frac{9}{10}}
  \exp\left(-\frac{2}{5}\frac{r}{r_{\mathrm{d}}(t)}\right)
  \,\mathrm{K}.
  \label{eq:temperature_vis}
\end{equation}
The corresponding disk scaleheight is calculated by
\begin{equation}
    H_{\mathrm{vis}} = 0.0337
    \left(\frac{r_{\mathrm{d}0}}{30\,\mathrm{au}}\right)^{-\frac{1}{5}}
    \left(\frac{\alpha}{10^{-3}}\right)^{\frac{1}{10}}
    \left(\frac{M_{\mathrm{d} 0}}{0.03\,M_{\odot}}\right)^{\frac{1}{5}}
    \left(\frac{t}{t_{\mathrm{diff}}}+1\right)^{-\frac{3}{10}}
    \left(\frac{r}{1\,\mathrm{au}}\right)^{\frac{21}{20}}
    \exp\left(-\frac{1}{5}\frac{r}{r_{\mathrm{d}}(t)}\right).
\end{equation}

Transforming Eq.~(\ref{eq:sigma_mdotin}), we obtain
\begin{equation}
  \Sigma_{\mathrm{g}} =
  \frac{\dot{M}_{\mathrm{in}}(t)}{3\pi \nu}
  \exp\left(-\frac{r}{r_{\mathrm{d}}(t)}\right)
  =
  \frac{\dot{M}_{\mathrm{in}}(t)}{3\pi \alpha H^{2}\Omega_{\mathrm{K}}}
  \exp\left(-\frac{r}{r_{\mathrm{d}}(t)}\right).
\end{equation}
Therefore, the surface density in the inner viscous heating region is given by
\begin{equation}
  \Sigma_{\mathrm{g,vis}} =
  605
  \left(\frac{r_{\mathrm{d}0}}{30\,\mathrm{au}}\right)^{-3/5}
  \left(\frac{\alpha}{10^{-3}}\right)^{-1/5}
  \left(\frac{M_{\mathrm{d} 0}}{0.03\,M_{\odot}}\right)^{3/5}
  \left(\frac{t}{t_{\mathrm{diff}}}+1\right)^{-3/10}
  \left(\frac{r}{1\,\mathrm{au}}\right)^{-3/5}
  \exp\left(-\frac{3}{5}\frac{r}{r_{\mathrm{d}}(t)}\right)
  \,\mathrm{g\,cm^{-2}}.
\end{equation}
In this model, the following criteria are used to connect the inner viscous-heating region
with the outer irradiation-heating region:
\begin{equation}
  T = \max\left(T_{\mathrm{vis}},\, T_{\mathrm{irr}}\right),
\end{equation}
\begin{equation}
  \Sigma_{\mathrm{g}} = \min\left(\Sigma_{\mathrm{g,vis}},\, \Sigma_{\mathrm{g,irr}}\right).
  \label{eq:select_sigma}
\end{equation}
We also note that, considering the reduction of opacity caused by the evaporation of silicate dust,
the temperature in the innermost region was fixed at 1400 K where $T$ exceeds 1400 K, for simplicity.

The definition in Eq.~(\ref{eq:select_sigma}) implies that the surface density in this model is necessarily
equal to or lower than that of the self-similar solution at every radius.
As a result, the total disk mass becomes smaller than $M_{\mathrm{d}}(t)$ defined by Eq.~(\ref{eq:mdself}),
which is based solely on the outer irradiated region.
In addition, as the viscous heating region shrinks
and its surface density approaches that of the self-similar solution,
the corresponding reduction in the “mass defect” leads to an artificial increase in the total disk mass,
resulting in an unphysical mass evolution.
Since the Monte Carlo simulation tracks particle motions based on the actual mass accretion rate distribution,
the simulation provides a more consistent and physically reliable evolution of the dust mass.
Therefore, for dust transport and distribution, the simulation outcome should be regarded as the appropriate reference.

In the simulation, we varied the turbulent viscosity parameter $\alpha$ ($10^{-2}$ and $10^{-3}$),
the initial disk mass parameter $M_{\mathrm{d0}}$ (0.01, 0.05, 0.1, and 0.2 $M_{\odot}$),
and the initial characteristic disk radius $r_{\mathrm{d0}}$ (1, 2, 5, 20, and 50 au),
resulting in 32 parameter combinations in total, excluding overly massive disks likely
to undergo gravitational instability \citep{Kimura2012Publ.Astron.Soc.Jpn., Kratter2016Annu.Rev.Astron.Astrophys.}.
Among these, the alpha values of $10^{-2}$ and $10^{-3}$ are adopted as representative values
of strong and weak turbulence, respectively, both supported by observational constraints
\citep{Flaherty2015ApJ, Flaherty2018ApJ} and MHD simulations \citep{Sano2004ApJ}.
Although even lower midplane turbulence levels ($\alpha \sim 10^{-4}$) have been inferred
from dust-settling and ring-structure observations \citep{Rosotti2023NewAstronomyReviews},
such low values are often interpreted as localized features in dead zones.
Therefore, $\alpha = 10^{-3}$ is adopted here as a conservative and broadly applicable choice
for weak turbulence, allowing for both physical relevance and computational feasibility.
In the self-similar framework, the disk mass parameter $M_{\mathrm{d0}}$ corresponds
to the total gas mass at the initial time, while the actual initial disk mass
in our combined viscous-irradiative disk model is somewhat lower as described above.
The initial characteristic radius $r_{\mathrm{d0}}$ is defined as the radius
at which the surface density falls to $1/e$ of its value at the inner edge of the disk.
However, this definition is purely theoretical and does not directly correspond
to observational disk sizes derived from $\mathrm{CO}$ emission flux \citep{Trapman2020A&A}
or centrifugal barrier constraints on disk edges \citep{Sakai2014Nature}.
Observationally inferred disk sizes span a wide range from a few au to over 100 au
and the chosen range of $r_{\mathrm{d0}}$ in this study is intended to cover this diversity.
These parameter sets are summarized in Table~\ref{tab:models_alpha-2}
and are consistently employed throughout this work.
In the model, the inner edge of the disk was set to be 0.1 au.
The initial accretion rate at the inner edge, $\dot{M}_{\mathrm{in}}(0)$,
can also be used as an alternative input parameter instead of $M_{\mathrm{d0}}$,
since they are physically linked through the viscous evolution.

\begin{table}
  \centering
  \caption{
    Model grid for $\alpha = 10^{-2}$ and $10^{-3}$.
    The same set of $(r_{\mathrm{d0}}, M_{\mathrm{d0}})$ combinations is adopted for both values of $\alpha$.
  }
  \label{tab:models_alpha-2}
  \begin{tabular}{c|ccccc}
    \hline
    $M_{\mathrm{d} 0}$ & \multicolumn{5}{c}{$r_{\mathrm{d}0}$} \\
    \cline{2-6}
     & 1 au & 2 au & 5 au & 20 au & 50 au \\
    \hline
    0.01$M_{\odot}$ & \included & \included & \included & \included & \included \\
    0.05$M_{\odot}$ & \included & \included & \included & \included & \included \\
    0.1$M_{\odot}$  & \included & \included & \included &           &           \\
    0.2$M_{\odot}$  & \included & \included & \included &           &           \\
    \hline
  \end{tabular}
\end{table}

\subsection{Trajectories of Dust Particles} \label{subsec:trajectory_model}
We conducted 3D Monte Carlo simulations to evaluate trajectories
of dust particles that move around by advection and diffusion, well coupled with disk gas
\citep{Ciesla2010ApJ, Ciesla2011ApJ, Okamoto2022ApJ, Okamoto2024A&A, Ishizaki2023ApJ, Yamamoto2024GeochimicaetCosmochimicaActa}.
The coordinate of a particle is tracked by the following equations:
\begin{equation}
  x_i = x_{i-1} + v_x \,\delta t_{\mathrm{phys}} + R_x \sqrt{6D(x')\,\delta t_{\mathrm{phys}}},
\end{equation}
\begin{equation}
  y_i = y_{i-1} + v_y \,\delta t_{\mathrm{phys}} + R_y \sqrt{6D(y')\,\delta t_{\mathrm{phys}}},
\end{equation}
\begin{equation}
  z_i = z_{i-1} + v_z \,\delta t_{\mathrm{phys}} + R_z \sqrt{6D(z')\,\delta t_{\mathrm{phys}}}.
\end{equation}
where $x$, $y$, and $z$ are the values of the Cartesian coordinates,
and $R_x$, $R_y$, and $R_z$ are independent random numbers $(-1 \leqq R \leqq 1)$.
The physical timestep $\delta t_{\mathrm{phys}}$ to calculate movement of dust particles
is given by $\delta t_{\mathrm{phys}} = \Omega_{\mathrm{K}}^{-1}$. Because of gradients in the diffusivity,
$x'$, $y'$, and $z'$ are given by $x' = x_{i-1} + (\partial D / \partial x)\delta t_{\mathrm{phys}}/2$,
$y' = y_{i-1} + (\partial D / \partial y)\delta t_{\mathrm{phys}}/2$
and $z' = z_{i-1} + (\partial D / \partial z)\delta t_{\mathrm{phys}}/2$.
The advection velocities in the viscous-irradiative combination disk are
\begin{equation}
  v_x = \left( v_r + \frac{1}{\Sigma_{\mathrm{g}}} \frac{\partial (D \Sigma_{\mathrm{g}})}{\partial r} \right) \frac{x}{r},
  \label{eq:pos_x}
\end{equation}
\begin{equation}
  v_y = \left( v_r + \frac{1}{\Sigma_{\mathrm{g}}} \frac{\partial (D \Sigma_{\mathrm{g}})}{\partial r} \right) \frac{y}{r},
  \label{eq:pos_y}
\end{equation}
\begin{equation}
  v_z = \frac{\partial D(z)}{\partial z} - \alpha \Omega_{\mathrm{K}} z.
  \label{eq:pos_z}
\end{equation}
where $r$ is the radial distance from the central star, given by $r = \sqrt{x^2 + y^2}$.
The radial drift velocity $v_{r}$ is given by Eq.~(\ref{eq:velo_r}),
meaning that dust particles move outward outside $r = r_{\mathrm{d}} (t) / 2$.
The second term in Eqs.~(\ref{eq:pos_x})--(\ref{eq:pos_z}) originates from radial gradient
of the accretion rate and vanishes in the case of a steady accretion disk
\citep{Ciesla2010ApJ, Ciesla2011ApJ, Okamoto2022ApJ, Okamoto2024A&A, Ishizaki2023ApJ, Yamamoto2024GeochimicaetCosmochimicaActa}.
For simplicity, the diffusivity of gas and dust, $D$, is replaced by the viscosity $\nu$.

\subsection{Progress of Irreversible Reactions} \label{subsec:reaction_model}
The progress of irreversible chemical reactions occurring in dust particles
as they migrate through a protoplanetary disk was calculated following the method of \citet{Ishizaki2023ApJ}.
These reactions were modeled using the Johnson-Mehl-Avrami (JMA) equation
\citep{Johnson1939Trans.AIME, Avrami1939J.Chem.Phys.}, which describes time-dependent reaction progress as:
\begin{equation}
  X = 1 - \exp\left[-\left(\frac{t}{\tau}\right)^n \right],
  \quad
  \tau = \left[ \nu_0 \exp\left(-\frac{E_{\mathrm{a}}}{RT}\right) \right]^{-1}.
\end{equation}
where $X$ is the degree of reaction progress ($0 \leqq X \leqq 1$),
$\tau$ is the characteristic timescale of the reaction, $t$ is time,
$n$ is the Avrami index, $\nu_{\mathrm{0}}$ is the pre-exponential factor, $E_{\mathrm{a}}$ is the activation energy,
$R$ is the gas constant. We collectively refer to the three quantities $n$, $\nu_0$,
and $E_{\mathrm{a}}$ as the “reaction parameters,” which differ depending on the specific reaction.
Each reaction adopts a unique set of these parameters. To numerically evaluate the reaction progress,
we integrated infinitesimal increments of the reaction degree $X$ using:
\begin{equation}
  \delta X =
  (1 - X)
  \left[
    1 - \exp\left(
      - n \frac{\delta t_{\mathrm{chem}}}{\tau}
      \left(-\ln(1 - X)\right)^{1 - \frac{1}{n}}
    \right)
  \right].
\end{equation}
where, $\delta t_{\mathrm{chem}}$ is the chemical timestep.

Because the reaction rate depends on the reaction degree,
we impose a cap on the progress per step: the maximum increment $\delta X_{\mathrm{max}}$.
The chemical timestep is dynamically adjusted at each step based on this cap:
\begin{equation}
  \delta t_{\mathrm{chem}} =
  \min\left(
    \delta t_{\mathrm{phys}},
    \delta t_{\mathrm{chem},\,\delta X = \delta X_{\max}}
  \right).
\end{equation}
where $\delta t_{\mathrm{phys}} = \Omega_{\mathrm{K}}^{-1}$ is the physical timestep
for Monte Carlo simulation,
and $\delta t_{\mathrm{chem},\delta X = \delta X{\mathrm{max}}}$ is the timescale
corresponding to $\delta X = \delta X_{\mathrm{max}}$, calculated as:
\begin{equation}
  \delta t_{\mathrm{chem},\,\delta X = \delta X_{\max}} =
  - \frac{\tau}{n \left(-\ln(1 - X)\right)^{1 - \frac{1}{n}}}
  \ln\left(
    1 - \frac{\delta X_{\max}}{1 - X}
  \right).
\end{equation}
When $\delta t_{\mathrm{chem},\delta X = \delta X{\mathrm{max}}}$ is selected,
the progress is advanced by $\delta X_{\mathrm{max}}$, and the particle trajectory is updated
assuming linear motion during this interval.
The updated condition is then used to recalculate $\delta X$ at the new location.
Without subdividing the timestep, a particle would complete the reaction using the reaction rate
evaluated at the beginning of the timestep,
even though the reaction rate changes continuously as $X$ increases.
In this study, we set $\delta X_{\mathrm{max}}=0.05$. Each reaction proceeds independently
and is considered complete when the reaction degree $X$ reaches 0.99.

We examined the progress of four representative irreversible reactions within a viscously evolving protoplanetary disk.
The reactions considered are: 
(1) crystallization of amorphous enstatite ($\mathrm{MgSiO_3}$) \citep{Kobayashi2023},
(2) crystallization of amorphous forsterite ($\mathrm{Mg_2SiO_4}$) \citep{Yamamoto2018ACSEarthSpaceChem.},
(3) crystallization of amorphous $\mathrm{MgFeSiO_4}$ \citep{Sakurai2024},
(4) thermal decomposition of kerogen \citep{Burnham1987EnergyFuels},
which cover effective temperatures between $\sim$500–1000 K in a steady accretion disk \citep{Ishizaki2023ApJ}.
We adopt kinetic parameters (Avrami index, pre-exponential factor,
and activation energy) derived from laboratory experiments.
All reaction parameters used in this study are summarized in Table~\ref{tab:irreversible_reactions}.
For each reaction, we recorded the maximum temperature experienced by every particle prior to reaction completion,
$T_{\mathrm{max}}$, which corresponds to the effective reaction temperature in a steady accretion disk \citep{Ishizaki2023ApJ}.

\begin{table}
  \centering
  \caption{
    Reaction parameters for the four reactions considered in this study,
    covering a temperature range of $\sim 500$--$1200\,\mathrm{K}$.
    Numbers in brackets refer to the sources:
    [1] \citet{Kobayashi2023},
    [2] \citet{Yamamoto2018ACSEarthSpaceChem.},
    [3] \citet{Sakurai2024},
    [4] \citet{Burnham1987EnergyFuels}.
  }
  \label{tab:irreversible_reactions}
  \begin{tabular}{lccc}
    \hline
    Reaction & $n$ & $E_a$ (kJ mol$^{-1}$) & $\ln[\nu_0\,(\mathrm{s}^{-1})]$ \\
    \hline
    $^{[1]}$Crystallization of amorphous En              & 1.5 & 850.0 & 81.0 \\
    $^{[2]}$Crystallization of amorphous Fo              & 1.5 & 414.4 & 40.2 \\
    $^{[3]}$Crystallization of amorphous MgFeSiO$_4$     & 1.25 & 323.9 & 34.9 \\
    $^{[4]}$Thermal decomposition of kerogen       & 1.0 & 231.1 & 32.8 \\
    \hline
  \end{tabular}
\end{table}

The crystallization kinetics of amorphous silicates are known to depend on their chemical composition,
initial structure, and surrounding environment during annealing.
In this study, we adopt the experimentally determined JMA kinetic parameters measured for amorphous silicates
synthesized by an induced thermal plasma method as representative examples.
We emphasize that the kinetic parameters we used in this study represent examples,
and that the model can readily incorporate other experimentally determined reaction parameters alternatively
\citep{Hallenbeck2000ApJ,Djouadi2005A&A,Murata2007ApJ},
provided that the reaction progress can be described by the general JMA equation.

Kerogen is used as a proxy material for complex organic solids, and its thermal decomposition is used
as a proxy for the destruction of refractory organic matter during thermal processing in the disk.
The degree of decomposition therefore serves as an indicator of carbon depletion in rocky materials.
The reaction kinetics adopted here \citep{Burnham1987EnergyFuels} was originally derived from laboratory experiments
on kerogen pyrolysis and therefore does not necessarily represent the thermal decomposition of refractory organic matter
in protoplanetary disks.
Nevertheless, the kinetics has been adopted in previous studies \citep{Li2021Sci.Adv.}
as an experimentally constrained description currently available.
Further experimental measurements of analog materials under disk-like conditions are required
to establish more appropriate reaction kinetics, and such experimental work is currently in progress.

These reactions were selected to span the wide temperature range relevant to dust processing in protoplanetary disks.
They also represent distinct classes of materials, including organics and compositionally diverse amorphous silicates,
which are found in primitive Solar System materials \citep{Yabuta2017GeochimicaetCosmochimicaActa}.
Taken together, these reactions provide a comprehensive basis for evaluating the thermal processing histories
of dust in protoplanetary disks, including the protosolar disk.

\section{Results} \label{sec:results}
\subsection{Predictive Formula for ``Reaction Line''} \label{subsec:reaction_line}
\citet{Ishizaki2023ApJ} showed detailed analysis of coupled dynamical and reaction evolution of individual particles and
developed a method to evaluate reaction progress of a particle swarm in steady accretion disks
without specifying chemical reactions, and showed that the ``reaction line'' temperature,
where a reaction proceeds efficiently, is determined by comparison
between the chemical reaction timescale and the timescale of dust diffusion.
The same approach can be extended to the combined viscous-irradiative disk.
However, unlike the time-independent steady accretion disk models,
the combined disk adopted in the present paper viscously expands and loses its mass with time.
To accommodate this, we apply the initial accretion rate at a representative radius $r_{\mathrm{T_{0}}}$,
defined as the location where the initial temperature equals the characteristic value $T_0$,
derived from reaction parameters alone (Eq.~\ref{eq:rough_temperature}).
This radius is numerically determined from Eqs.~(\ref{eq:temperature_irr}) and (\ref{eq:temperature_vis}),
and the corresponding accretion rate is calculated by substituting $r_{\mathrm{T_{0}}}$
into the mass accretion rate profile of the combined disk model using Eq.~(\ref{eq:mdot}):
\begin{equation}
  \dot{M}_0(r) =
  \dot{M}_{\mathrm{in}}(0)
  \left[
    1 - \left(\frac{r}{r_{\mathrm{d}}(t)/2}\right)
  \right]
  \exp\left(-\frac{r}{r_{\mathrm{d}}(t)}\right).
\end{equation}
This estimated accretion rate is then used in the predictive formula Eq.~(\ref{eq:predicted_line_temperature}),
enabling its application to the combined disk framework.
Note that the radius corresponding to $T_0$ is based on the initial temperature distribution,
as the temperature of the reaction line remains nearly unchanged over the simulation timescale.

In certain cases such as low-temperature reactions occurring in initially massive, compact disks,
the corresponding accretion rate can become close to zero or even negative (i.e., outward advection).
Since the predictive formula Eq.~(\ref{eq:predicted_line_temperature}) assumes a monotonic,
inward accretion flow linked to the temperature structure, such regions fall outside its original scope.
To maintain consistency in our analysis, we set a minimum value of $\dot{M} = 10^{-9}M_\odot / \mathrm{yr}$
whenever the calculated accretion rate falls below this threshold.
This threshold is chosen for both physical and practical reasons:
first, the predicted reaction line temperature shows only weak sensitivity to the accretion rate,
so the impact of fixing $\dot{M}$ is minor;
second, $\dot{M} = 10^{-9}M_\odot / \mathrm{yr}$ represents a plausible lower bound
for typical disk accretion rates \citep{Alexander2023Mon.Not.R.Astron.Soc.},
ensuring that the adopted value remains within a realistic regime.

\begin{figure}
\begin{center}
\includegraphics[width=10cm,clip]{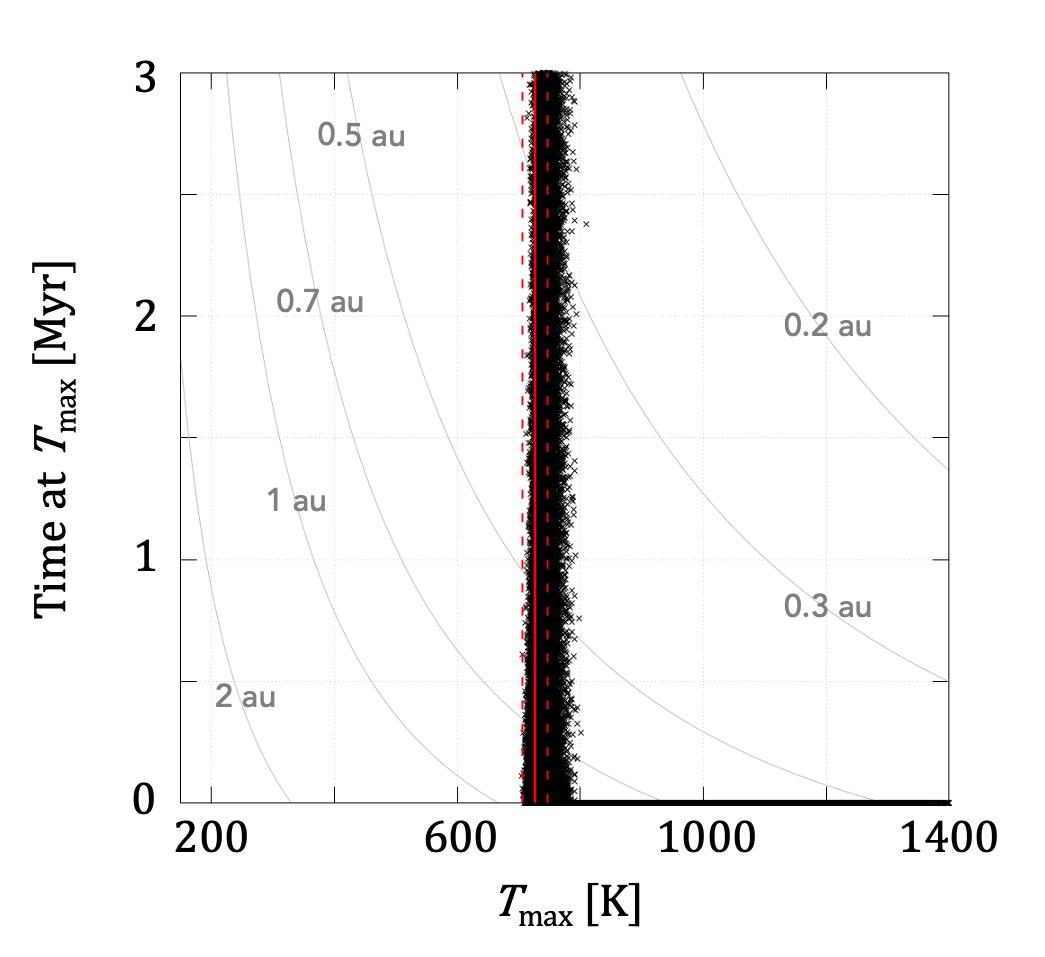}
  \caption{Time and radial locations at which dust particles reached their maximum temperature ($T_\mathrm{max}$)
    prior to completion of crystallization of amorphous forsterite.
    Each dot represents a dust particle, plotted at the location and time of $T_\mathrm{max}$
    (model: $\alpha = 10^{-3}$, $r_{\mathrm{d0}} = 5 \, \mathrm{au}$, $M_{\mathrm{d0}} = 0.05 M_{\odot}$).
    Predicted reaction lines are overlaid: solid red lines indicate the reaction line temperatures,
    while dashed red lines show the corresponding dispersions,
    both calculated from the analytical formula
    (Eqs.~(\ref{eq:predicted_line_temperature}) and (\ref{eq:predicted_line_width})).
  }
  \label{fig:reaction_line}
  \end{center}
\end{figure}

Figure~\ref{fig:reaction_line} shows $T_{\mathrm{max}}$ and the corresponding time for each dust particle,
together with the predicted reaction line (red line).
A solid line represents the reaction line temperature calculated from Eq.~(\ref{eq:predicted_line_temperature}),
while dashed lines denote the expected dispersion derived from Eq.~(\ref{eq:predicted_line_width}).
The predicted lines show good agreement with the numerical results.
Across a range of disk parameters and reaction types, the predicted lines consistently align
with the maximum temperatures experienced by dust particles prior to reaction completion
(cf. Figures~\ref{fig:representative_colormap}, \ref{fig:comet_colormap} and \ref{fig:additional_colormap}).
The deviations remain within acceptable ranges, given the inherent ambiguity in defining a precise reaction line.
This formulation provides a useful framework for understanding the spatial distribution of dust
that has undergone distinct chemical processes in protoplanetary disks.

\subsection{Radial Distribution of Reacted Dust Particles} \label{subsec:radial_distribution}
Dust particles that undergo irreversible chemical reactions are redistributed throughout the disk
via advective and diffusive transport, leading to complex patterns of spatiotemporal structure.
The efficiency of these processes strongly depends on disk parameters.

\begin{figure}
\begin{center}
  \includegraphics[
  width=\textwidth,
  height=0.9\textheight,
  keepaspectratio
  ]{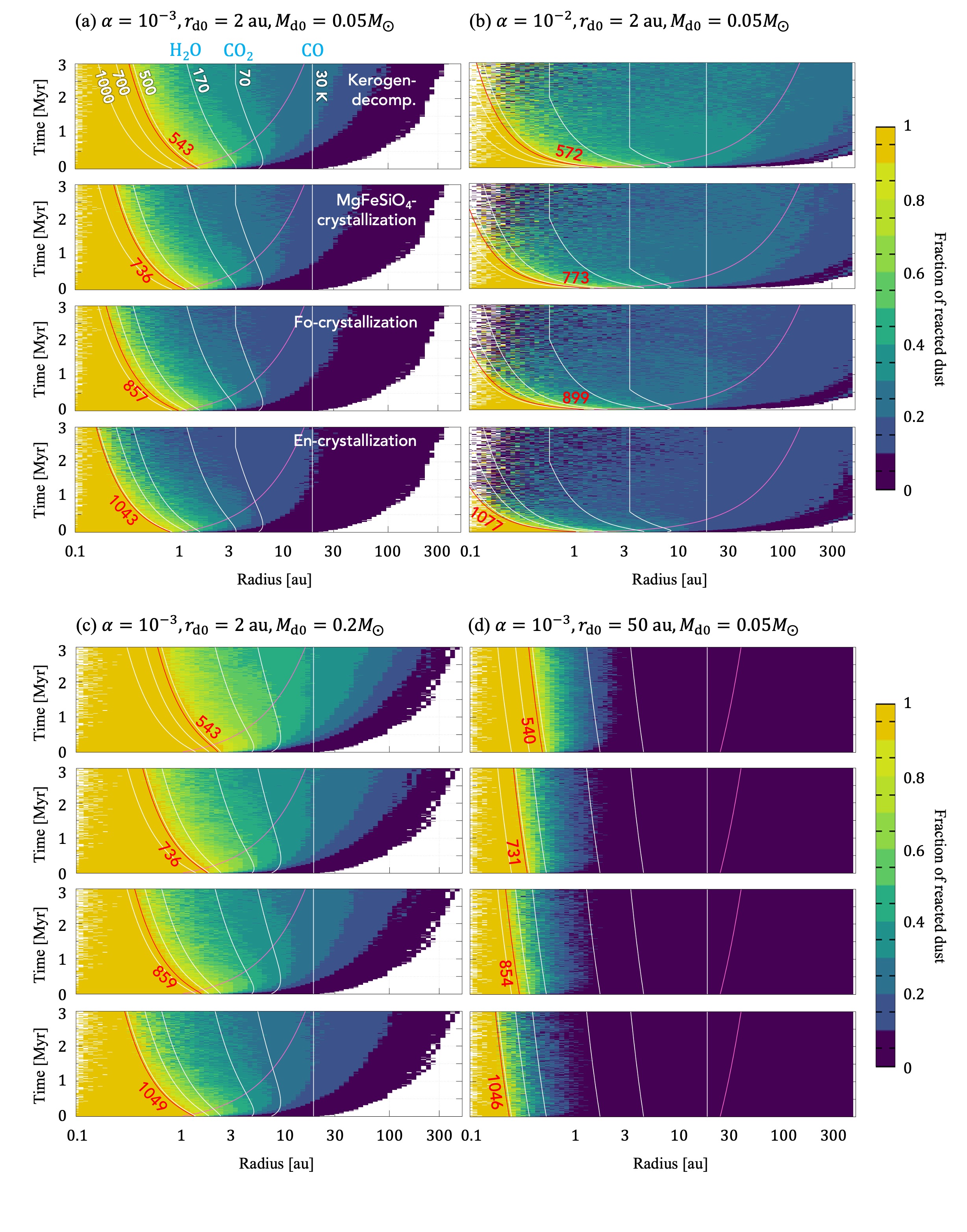}
  \caption{
    Time evolution of the radial distribution of dust that has completed each chemical reaction considered in this paper.
    Each panel shows the fraction of reacted dust particles in radius–time space,
    for a given combination of disk parameter and reaction type. White contour lines represent isotherms,
    and red and pink lines indicate the reaction line and the stagnation line, respectively.
    Kerogen-decomp.: thermal decomposition of kerogen;
    Fo-crystallization: crystallization of amorphous forsterite (Mg$_2$SiO$_4$));
    MgFeSiO$_4$-crystallization: crystallization of amorphous MgFeSiO$_4$;
    En-crystallization: crystallization of amorphous enstatite (MgSiO$_3$).
    Each column shows four models, as indicated by the labels at the top.
    }
  \label{fig:representative_colormap}
  \end{center}
\end{figure}

Figure~\ref{fig:representative_colormap} illustrates the time evolution of the radial distribution of dust
that has completed each chemical reaction examined in this study.
Although each grid cell in the plot displays the mean reaction degree of all particles it contains,
the values closely approximate the fraction of reacted dust because the system exhibits a near-binary behavior
-- most particles are either fully reacted or entirely unreacted.

The four-row panels in each column of Figure~\ref{fig:representative_colormap} display the spatiotemporal distribution
of reacted dust for each chemical process, arranged from top to bottom in order of decreasing reaction temperature.
As expected, lower-temperature reactions (upper panels) yield broader distributions of reacted dust,
while higher-temperature reactions (lower panels) are more spatially confined. Regarding disk parameters,
the trends are summarized as following:
(1) turbulent disks with larger $\alpha$ show broader spatial distributions of reacted dust due to enhanced radial mixing
(compare Figure~\ref{fig:representative_colormap} (a) and (b));
(2) more massive disks produce greater amounts of reacted dust as a result of elevated temperatures across the disk
(compare Figure~\ref{fig:representative_colormap} (a) and (c));
(3) compact disks facilitate both efficient reaction and outward transport due to the large hot region and extended advective outflow zone
(compare Figure~\ref{fig:representative_colormap} (a) and (d)).

Despite differences in reaction types, the spatial distribution of reacted dust often exhibits similar patterns
within the same disk model. In particular, whether the reacted dust is transported beyond a certain key radius,
such as the H$_2$O snowline, tends to depend more on disk dynamics than on reaction type.
However, notable divergences arise when the reactions require markedly different temperatures.
For instance, in a model with $\alpha = 10^{-2}$, $r_{\mathrm{d0}} = 5 \, \mathrm{au}$, and $M_{\mathrm{d0}} = 0.05 M_{\odot}$
(Figure~\ref{fig:additional_colormap}(a)), dust undergoing lower-temperature reactions is efficiently transported outward,
whereas dust that experienced high-temperature reactions remains trapped near the innermost region.
This discrepancy occurs because the high-temperature reactions are confined to a narrower inner zone,
limiting both the spatial and temporal window for reaction and subsequent outward drift.

Another noteworthy feature appears in some of the highly turbulent disk models:
a population of reacted dust remains in the outer disk after 3 Myr,
while the reacted dust in the inner disk accretes onto the central star.
For instance, in a model with $\alpha = 10^{-2}$, $r_{\mathrm{d0}} = 2 \, \mathrm{au}$, and $M_{\mathrm{d0}} = 0.1 M_{\odot}$
(Figure~\ref{fig:additional_colormap}(b)) after $\sim$2–3 Myr, the crystalline enstatite fraction is low (0-0.1)
across many radial bins within 0.1-0.5 au, whereas it reaches moderate levels (0.1-0.3) in ~1-10 au.
In these models, the hot inner region -- necessary for progress of chemical reactions
-- disappears within a few Myr due to rapid cooling, in contrast to more quiescent disks
where high-temperature regions persist longer. When combined with efficient outward transport at early stages,
this leads to a counterintuitive outcome: while reacted dust remaining in the inner region is gradually lost
by accretion onto the central star,
a fossil population of reacted grains survives in the outer disk, having been exported before the inner region cooled.
This spatial decoupling between processed and unprocessed dust is not seen in steady-state disks,
highlighting the unique dynamical-chemical imprint of viscously evolving disks.

\subsection{Distribution Pattern: ``Reaction Line'' and ``Stagnation Line''} \label{subsec:distribution_pattern}
Distributions of dust that experience chemical reactions are fundamentally controlled by two processes:
(1) the location where reactions proceed effectively and
(2) the transport of processed dust (redistribution).
These processes are dictated by two spatial boundaries:
the reaction line, a narrow temperature-defined region where a given reaction completes efficiently,
and the stagnation line, the radial location where advection reverses direction.
The location of the stagnation line ($r_{\mathrm{stag}}$) in this model is derived
from the radial velocity profile Eq.~(\ref{eq:velo_r}). It is analytically given by half of the disk radius at each time,
\begin{equation}
  r_{\mathrm{stag}} = \frac{r_d(t)}{2}.
\end{equation}
Dust located beyond the stagnation line is preferentially transported outward,
while dust inside moves inward and accretes onto the central star.
The relative positioning of these lines therefore governs whether processed dust
is retained in the inner disk or exported to outer regions.

Figure~\ref{fig:representative_colormap} overlays reaction lines (red) and stagnation lines (pink) on maps of reacted dust fractions.
In all models, regions interior to the reaction line consistently show the highest levels of reacted dust,
visibly corresponding to yellow zones, indicating that the reaction line serves as a robust outer boundary
for active chemical processing.
In some cases where the reaction line initially lies somewhat beyond the stagnation line
(e.g., all reactions in Figure~\ref{fig:representative_colormap}(c)),
the boundary of the high-reacted fraction extends slightly farther outward,
reflecting early-stage transport of reacted grains.
These models exhibit high reacted fractions across much of the disk, with few completely unreacted (dark blue) regions.

\begin{figure}
  \begin{center}
    \includegraphics[width=0.65\textwidth]{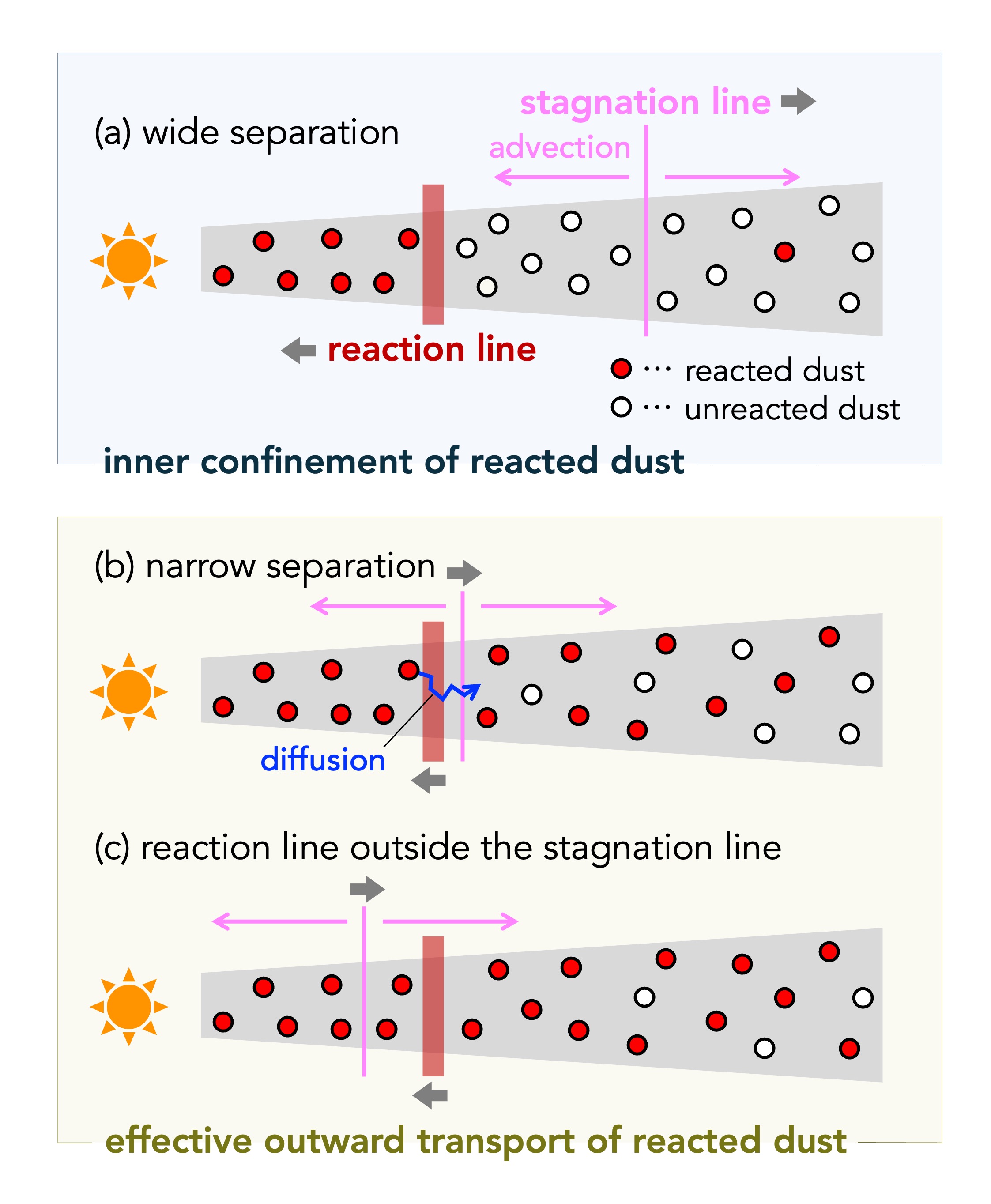}
    \caption{
      Classification scheme based on the relative positions of the two lines:
      (a) a wide separation between the lines;
      (b) a narrow separation between the two lines;
      and (c) the reaction line located outside the stagnation line.
      Case (a) results in inner confinement of reacted dust,
      whereas cases (b) and (c) result in efficient outward transport of reacted dust.
    }
    \label{fig:classification}
  \end{center}
\end{figure}

Based on the spatial relationship between the reaction and stagnation lines,
we identify two characteristic classes of distribution patterns.
A schematic summary is shown in Figure~\ref{fig:classification},
which classifies the distributions according to the relative positions of the two lines.
The first group exhibits a wide separation between the two lines and includes models
with large initial disk radii ($r_0 > 20 \, \mathrm{au}$; e.g., Figure~\ref{fig:representative_colormap}(d)).
In these cases, the reaction line lies deep within the inner disk,
where the accretion rate remains relatively constant.
As a result, only a small fraction of reacted dust diffuses outward,
and even fewer grains reach the stagnation line.
The processed dust is thus confined to the inner disk,
yielding a compact distribution similar to steady-state accretion models.

In the second group, the reaction line is located near or beyond the stagnation line
at the beginning of the calculation.
This group is mainly associated with smaller initial disks ($r_0 < 5 , \mathrm{au}$;
e.g., Figure~\ref{fig:representative_colormap}(a)-(c)).
From the reaction line, a gradient emerges as reacted dust diffuses outward beyond the stagnation line,
forming intermediate bands distinct from both the highly reacted inner zones (yellow)
and the unreacted outer disk (dark blue).
These bands appear as color gradients or “reaction fronts”
in Figure~\ref{fig:representative_colormap}, \ref{fig:comet_colormap} and \ref{fig:additional_colormap}
-- visual markers of how local transport reshapes the distribution of processed dust:
(1) inside the stagnation line, the fronts curve along the reaction line, shaped by inward advection;
(2) near the stagnation line, the fronts stand nearly vertical, indicating the transition in flow direction;
(3) beyond the stagnation line, the fronts follow the contour of the stagnation line or disk edge,
maintained by outward advection. This zonation reflects the kinematic structure of the disk.

These results demonstrate that the spatial pattern of chemical processing is dictated
by the interplay between reaction progress and dust transport.
The relative positioning of the reaction line and the stagnation line acts as a structural control,
determining both where reactions can proceed and whether processed dust is retained or exported.
Their configuration directly shapes the final distribution of reaction products in the disk.

\subsection{Crystallinity in the Solar System Materials} \label{subsec:solar_system_crystallinity}
Comets exhibit crystalline silicate fractions ranging from 10 to 60\%
\citep{Harker2011TheAstronomicalJournal, Sitko2011TheAstronomicalJournal, Shinnaka2018AJ},
along with diversity in the olivine-to-pyroxene ratio \citep{Shinnaka2018AJ}. 
These observations imply that crystalline silicate dust must have been transported outward
after crystallization from the amorphous state in the hot inner disk, which requires high temperatures.
Here, we examine which disk models can reproduce the observed crystalline silicate properties of the Solar System.

Previous studies have shown that turbulent diffusion alone produces a crystalline fraction of only $\sim$5\% in the comet forming region,
and that additional processes are required to account for the crystallinity observed in comets \citep{Okamoto2022ApJ}. 
In contrast, our results show that crystalline silicate dust can be efficiently transported outward
when the crystallization line is located near or beyond the stagnation line, as a natural consequence of disk evolution.
This behavior is not captured in steady disk models, but can arise naturally when an initially compact disk is considered.

\begin{figure}
\begin{center}
  \includegraphics[
  width=\textwidth,
  height=0.9\textheight,
  keepaspectratio
  ]{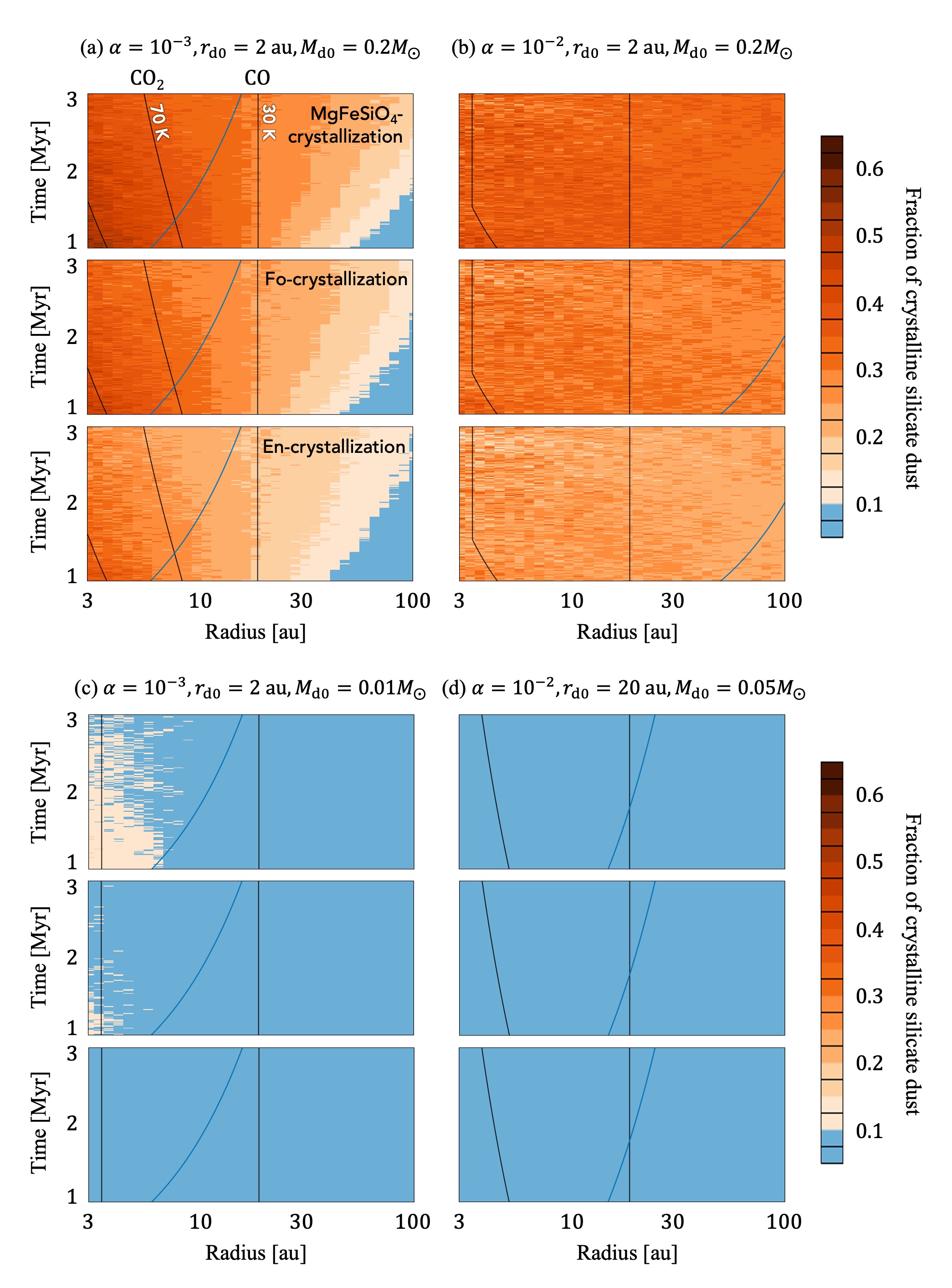}
  \caption{Time evolution of the radial distribution of the fraction of crystalline silicate dust, 
    focusing on the comet-forming region ($r = 3$–$100 \, \mathrm{au}$ and $t = 1$–$3 \, \mathrm{Myr}$). 
    White contour lines represent isotherms, and the blue line indicates the stagnation line. 
    Colors show the degree of crystallization, where values below 0.1 are grouped into a single color for clarity,
    intermediate values (0.1–0.6) are resolved in finer bins than
    Figure~\ref{fig:representative_colormap}, and higher values are saturated. 
    Fo-crystallization: crystallization of amorphous forsterite (Mg$_2$SiO$_4$); 
    MgFeSiO$_4$-crystallization: crystallization of amorphous MgFeSiO$_4$; 
    En-crystallization: crystallization of amorphous enstatite (MgSiO$_3$).
    Each column shows four models, as indicated by the labels at the top.
    }
  \label{fig:crystallinity}
  \end{center}
\end{figure}

Figure~\ref{fig:crystallinity} shows the radial distribution of the degree of crystallization,
focusing on the comet forming region.
The region between the CO$_2$ snowline and 100 au from $t = 1$ to $3 \,\mathrm{Myr}$ is adopted as the comet forming region.
We do not assume a specific comet formation scenario, because our objective is to evaluate the availability
and spatial variation of crystalline silicates throughout the broad spatiotemporal region where comets may form
under various proposed formation scenarios \citep{Weidenschilling1997Icarus,Blum2022Universe}.
The conclusions are based on the overall distribution of crystalline silicates
rather than on a particular comet formation scenario (e.g., the formation location, timing, and duration),
and are therefore expected to be robust to different assumptions regarding comet formation.

Compact disks ($r_0 < 5 \,\mathrm{au}$) exhibit higher crystallinity in the outer disk,
often reaching 10–60\% (Figure~\ref{fig:crystallinity}(a)),
comparable to observations.
This is because the crystallization lines (736 K for MgFeSiO$_4$, 859 K for forsterite, and 1049 K for enstatite
in the model with $\alpha=10^{-3}$, $r_0=2\,\mathrm{au}$ and $M_0=0.2\,M_\odot$)
are initially located beyond the stagnation lines, allowing efficient outward transport of crystalline silicates
(Figure~\ref{fig:comet_colormap}(a)).
For $\alpha=10^{-2}$ (Figure~\ref{fig:crystallinity}(b)), the crystallization lines are also located beyond the stagnation lines
(Figure~\ref{fig:comet_colormap}(d); 773 K for MgFeSiO$_4$, 899 K for forsterite, and 1077 K for enstatite
in the model with $\alpha=10^{-2}$, $r_0=2\,\mathrm{au}$ and $M_0=0.2\,M_\odot$),
leading to efficient redistribution of crystalline dust.
However, the crystalline fraction remains nearly constant across the comet-forming region for each composition.
This behavior is caused by efficient mixing due to strong turbulence.
Therefore, highly turbulent disks are not favorable for reproducing the observed diversity in the olivine-to-pyroxene ratio.

Additionally, very low-mass disks ($M_0 = 0.01\,M_{\odot}$; Figure~\ref{fig:crystallinity}(c)) fail to reach sufficiently high temperatures,
particularly for crystallization of amorphous enstatite, even with initially compact disk sizes.
In these cases, the lower disk temperature places the crystallization lines
(730 K for MgFeSiO$_4$, 852 K for forsterite, and 1044 K for enstatite
in the model with $\alpha=10^{-3}$, $r_0=2\,\mathrm{au}$ and $M_0=0.01\,M_\odot$) close to the central star,
resulting in a wide separation between the stagnation line and the crystallization line (Figure~\ref{fig:comet_colormap}(c)).
Such models are therefore unfavorable for reproducing both the crystallinity and the diversity in the olivine-to-pyroxene ratio observed in comets.

In contrast, disks with relatively large initial sizes ($r_0 > 20 \,\mathrm{au}$) show limited outward transport of crystalline silicates,
including MgFeSiO$_4$, which has the lowest crystallization temperature among the compositions considered here
(Figure~\ref{fig:crystallinity}(d)).
In these models, the crystallization lines (758 K for MgFeSiO$_4$, 883 K for forsterite, and 1066 K for enstatite
in the model with $\alpha=10^{-2}$, $r_0=20\,\mathrm{au}$ and $M_0=0.05\,M_\odot$)
remain well inside the stagnation line (Figure~\ref{fig:comet_colormap}(d)),
because the stagnation line is located at large radii from the beginning.

These results suggest that the protosolar disk was likely compact ($r_0 < 5 \,\mathrm{au}$),
moderately massive ($M_0 > 0.05\,M_{\odot}$), and not strongly turbulent to account for crystalline silicate dust in comets.

 Beyond their implications for the protosolar disk, these results also provide testable predictions
 for observations of protoplanetary disks.
 They predict the spatiotemporal distributions of dust that has undergone crystallization
 or other irreversible reactions under different disk conditions.
 Coupling these distributions with radiative transfer calculations will enable direct comparisons
 with spatially resolved mid-infrared observations of protoplanetary disks,
 providing observational constraints on their physical properties and evolution.
 Although the present study focuses on dust, the same framework can be extended to gas-phase species.
 Gas-phase molecules released by thermal decomposition of refractory organic matter originate in the same hot inner-disk region
 as thermally processed solids and may therefore exhibit correlated spatial distributions \citep{Nuth2000Nature}.
 A quantitative comparison with observations would require additional processes,
 such as condensation beyond the snowlines of individual species, to be taken into account,
 because condensation changes the size and transport properties of particles.

\section{Conclusions} \label{sec:conclusions}
Irreversible chemical reactions leave observable imprints in Solar System materials,
providing powerful constraints on the early evolution of the protosolar disk.
In this study, we examined how such reactions proceed in a viscously evolving protoplanetary disk
and how the spatial distribution of reacted dust evolves over time, using Monte Carlo simulations.
We developed a combined viscous-irradiative disk model that enables a thermally and dynamically consistent treatment of evolving disks,
including accurate temperature distributions in the hot inner region.

Across a broad range of disk parameters and reaction types, each irreversible reaction proceeds within a narrow temperature range,
allowing the definition of a characteristic reaction line temperature for each reaction.
The predictive formula developed for steady accretion disks can be extended to evolving viscous–irradiative disks
by evaluating the accretion rate at the characteristic reaction temperature.
This generalization enables prediction of chemically processed dust distributions from reaction kinetics,
providing a versatile diagnostic framework.

The distribution of reacted dust reflects the combined effects of reaction kinetics and disk dynamics.
Compact, massive, and turbulent disks promote efficient formation and outward transport of processed grains.
The spatial extent of the reaction zone also plays a key role:
high-temperature reactions are restricted to the innermost disk, whereas lower-temperature reactions occur over broader regions,
allowing processed grains to reach larger radii.
In highly turbulent disks, the outer disk can eventually retain more reacted dust than the inner disk,
a counterintuitive outcome arising from rapid early outward transport and subsequent cooling of the inner region.

The spatial pattern of processed dust is fundamentally governed by the relative positioning of the reaction line and the stagnation line,
which marks the boundary between inward and outward advection.
When these lines are widely separated, reacted dust remains confined to the inner disk;
when the reaction line locates near or beyond the stagnation line, processed dust is redistributed outward even in the region
where temperatures never become high enough for effective reactions.
These results demonstrate that the interplay between reaction progress
and global dust transport determines the chemical structure of the disk.

Finally, we assessed the disk conditions most consistent with the crystalline silicate observed in comets.
Unlike previous studies that invoked additional transport processes,
our model achieves efficient outward transport of crystalline silicates as a natural consequence of disk evolution.
The results suggest that the protosolar disk was likely compact ($r_{\mathrm{d} 0} < 5 \,\mathrm{au}$),
moderately massive ($M_{\mathrm{d} 0} > 0.05\,M_{\odot}$), and not strongly turbulent.

The present framework also provides a basis for direct comparison
with spatially resolved observations through radiative transfer calculations
and for future extensions to gas-phase chemistry,
  enabling broader observational constraints on the evolution of protoplanetary disks.

\begin{acknowledgments}
  We are grateful to the anonymous referee for the careful reading of the manuscript
  and many thoughtful and constructive comments, which significantly improved the quality and clarity of this work.
  This work is supported by the KAKENHI grant 25KJ0913, 20H05844 and 24H00265.
  The authors acknowledge the use of ChatGPT (OpenAI) for manuscript editing and language refinement.
  All authors have reviewed and approved the final version of the text.
\end{acknowledgments}

\appendix
\section{Predictive Formula for ``Reaction Line'' in a Steady Accretion Disk} \label{app:predictive_formula}
\citet{Ishizaki2023ApJ} demonstrated that in a steady accretion disk,
the temperature at which a dust particle completes an irreversible reaction (``reaction line'' temperature)
as well as its dispersion, can be quantitatively predicted.
The formulation is based on comparing the chemical reaction timescale to the timescale of dust diffusion.
This framework enables an assessment of whether dust particles will complete a given reaction,
without requiring explicit numerical integration of the reaction kinetics.

The predictive formula integrates both reaction parameters (the Avrami index $n$,
the pre-exponential factor $\nu_{\mathrm{0}}$, the activation energy $E_{\mathrm{a}}$,
and the required degree of reaction $X$) and disk parameters
(the dimensionless viscosity $\alpha$ and the accretion rate $\dot{M}$):
\begin{equation}
  T_{\mathrm{reac}} =
  \frac{E_a}{R}
  \Biggl[
  16.15 - \ln C_X + \ln\!\bigl(\nu_{\mathrm{0}}\, [\mathrm{s}^{-1}]\bigr)
  + 2 \ln\!\left(\frac{\sigma_{\mathrm{reac}}}{0.0162}\right)
  - \frac{14}{9} \ln\!\left(\frac{T_0}{10^3\,\mathrm{K}}\right)
  - \frac{10}{9} \ln\!\left(\frac{\alpha}{10^{-2}}\right)
  + \frac{2}{9} \ln\!\left(\frac{\dot{M}}{10^{-7}\,M_\odot\,\mathrm{yr}^{-1}}\right)
  \Biggr]^{-1}
  \,\mathrm{K}.
  \label{eq:predicted_line_temperature}
\end{equation}
Here, $C_X$ is a parameter related to the reaction defined as:
\begin{equation}
  C_X =
  \frac{1}{n}
  \left[-\ln(1 - X)\right]^{-\frac{n-1}{n}}
  \qquad (n < 1),
\end{equation}
\begin{equation}
  C_X = \frac{1}{n}
  \qquad (n \ge 1),
\end{equation}
and $T_0$ is a characteristic temperature representing a rough estimate of the reaction line,
determined solely from the reaction kinetics:
\begin{equation}
  T_0 =
  \frac{E_a}{R}
  \left[
    16.15 - \ln C_X + \ln\!\bigl(\nu_0\, [\mathrm{s}^{-1}]\bigr)
  \right]^{-1}
  \,\mathrm{K}.
  \label{eq:rough_temperature}
\end{equation}
The relative dispersion of the reaction line ($\sigma_{\mathrm{reac}}$) is given by:
\begin{equation}
  \sigma_{\mathrm{reac}} =
  0.72 \times
  \left[
    16.15 - \ln C_X + \ln\!\bigl(\nu_{\mathrm{0}}\, [\mathrm{s}^{-1}]\bigr)
  \right]^{-1}.
  \label{eq:predicted_line_width}
\end{equation}
This formulation successfully reproduces the reaction progress of a wide range of irreversible reactions
(with reaction temperatures spanning $\sim200-1400 \, \mathrm{K}$) in a steady accretion disk.
Its broad applicability makes it a powerful tool for bridging chemical kinetics with disk dynamics,
especially in regimes where direct simulations are computationally expensive.

\section{Additional Examples of Reacted Dust Distributions} \label{app:color_maps}
This appendix presents additional examples of reacted dust distributions
that illustrate characteristic behaviors discussed in Section~\ref{subsec:distribution_pattern}.
These examples highlight (i) differences in spatial distributions among reaction types within the same disk model
and (ii) the survival of a fossil population of reacted dust in highly turbulent disks.

\begin{figure}
\begin{center}
  \includegraphics[
  width=\textwidth,
  height=0.9\textheight,
  keepaspectratio
  ]{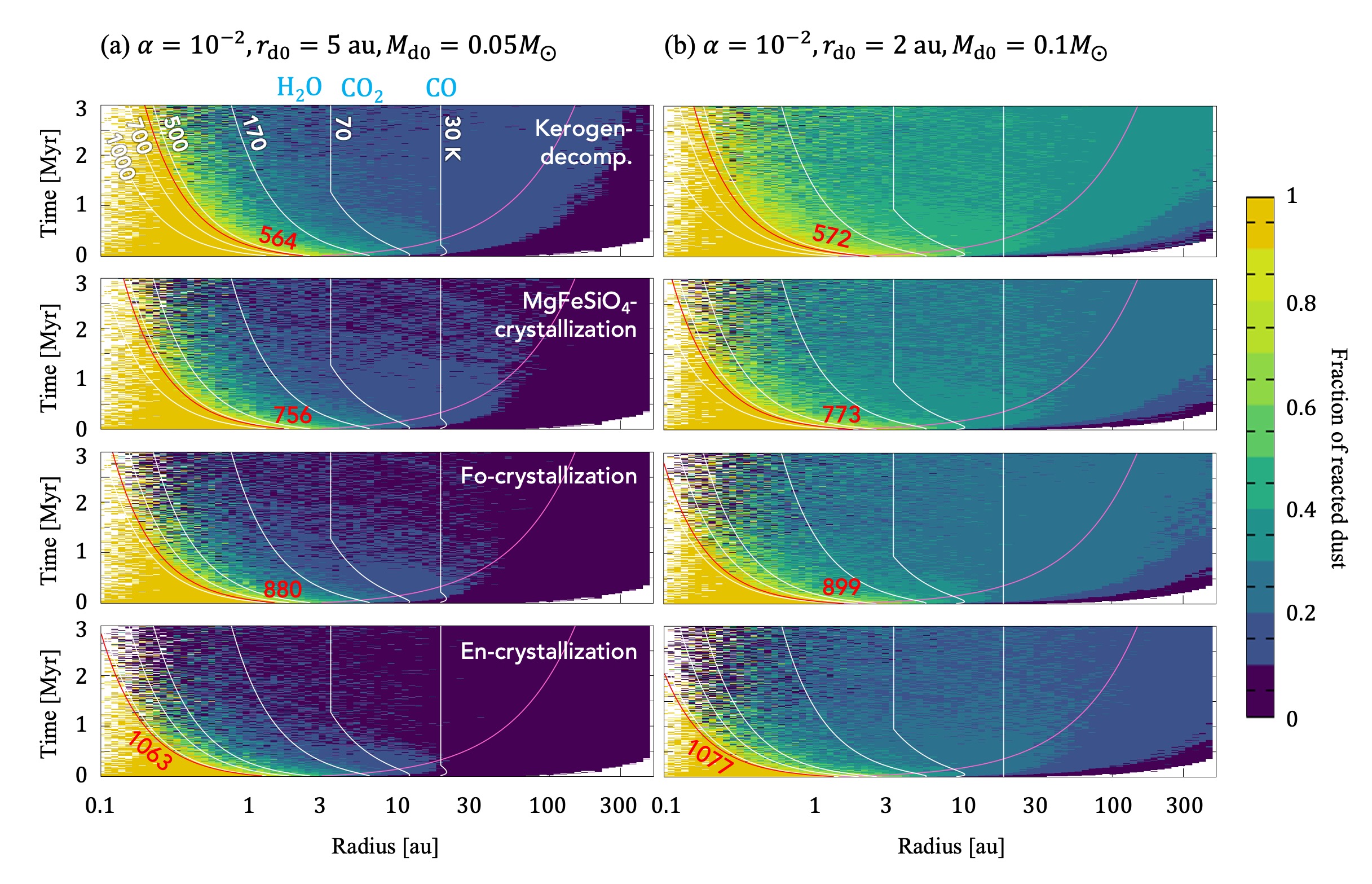}
  \caption{
    Additional examples of Figure~\ref{fig:representative_colormap}.
    Time evolution of the radial distribution of dust that has completed each chemical reaction considered in this paper.
    Each panel shows the fraction of reacted dust particles in radius--time space
    for a given combination of disk parameters and reaction type.
    White contour lines represent isotherms,
    while red and pink lines indicate the reaction line and the stagnation line, respectively.
  }
  \label{fig:additional_colormap}
  \end{center}
\end{figure}

Figure~\ref{fig:additional_colormap}(a) illustrates that dust undergoing high-temperature reactions
(e.g., crystallization of amorphous enstatite) is confined to the inner disk,
whereas dust undergoing lower-temperature reactions (e.g., thermal decomposition of kerogen)
is transported efficiently outward in the model
with $\alpha = 10^{-2}$, $r_{\mathrm{d0}} = 5 \, \mathrm{au}$, and $M_{\mathrm{d0}} = 0.05 M_{\odot}$.

Figure~\ref{fig:additional_colormap}(b) shows a model with $\alpha = 10^{-2}$, $r_{\mathrm{d0}} = 2 \, \mathrm{au}$,
and $M_{\mathrm{d0}} = 0.1 M_{\odot}$, in which crystalline enstatite remains at a higher fraction
in the outer disk than in the inner disk.
As a result, crystalline dust remaining in the inner disk is gradually lost through accretion onto the central star,
whereas grains transported outward at earlier times remain preserved in the outer disk.

\section{Full Distributions of Crystalline Silicates} \label{app:comet_colormaps}
\begin{figure}
\begin{center}
  \includegraphics[
  width=\textwidth,
  height=0.9\textheight,
  keepaspectratio
  ]{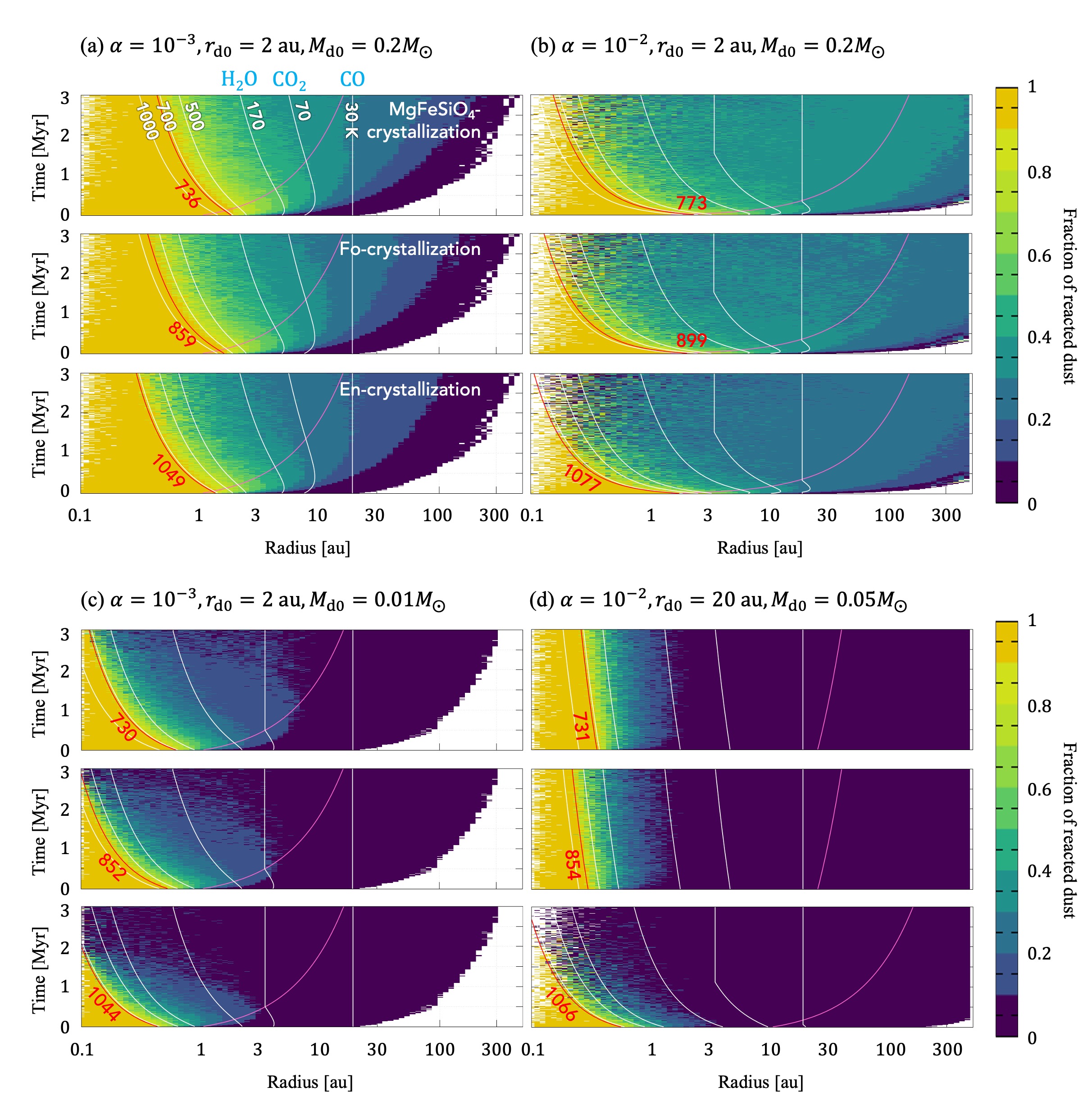}
  \caption{
    Extended view of Figure~\ref{fig:crystallinity}.
    The same data are shown over the full spatial and temporal range of the simulation.
  }
  \label{fig:comet_colormap}
  \end{center}
\end{figure}

Figure~\ref{fig:comet_colormap} shows the same crystallinity distributions as Figure~\ref{fig:crystallinity},
but over the full spatial and temporal range of the simulation.

\bibliography{mybib}{}
\bibliographystyle{aasjournalv7}
\end{document}